# Enhancing magnetocrystalline anisotropy of the $Fe_{70}Pd_{30}$ magnetic shape memory alloy by adding Cu

S Kauffmann-Weiss[1,2], S Hamann[3], M E Gruner[4], L Schultz[1,2], A Ludwig[3] and S Fähler[1]

[1] IFW Dresden, P.O. Box 270116, 01171 Dresden, Germany

[2] Dresden University of Technology, Institute for Materials Science, 01062 Dresden, Germany

[3] Ruhr-Universität Bochum, Institute for Materials, 44780 Bochum, Germany

[4] University of Duisburg-Essen, Theoretical Physics, 47048 Duisburg, Germany

E-mail: s.weiss@ifw-dresden.de

**Abstract.** Strained epitaxial growth provides the opportunity to understand the dependence of intrinsic and extrinsic properties of functional materials at frozen intermediate stages of a phase transformation. In this study, a combination of thin film experiments and first-principles calculations yields the binding energy and magnetic properties of tetragonal $Fe_{70}Pd_{30-x}Cu_x$ ferromagnetic shape memory thin films with x = 0, 3, 7 and structures ranging from *bcc* to beyond *fcc* ($1.07 < c/a_{bct} < 1.57$).

We find that Cu enhances the quality of epitaxial growth, while spontaneous polarisation and Curie temperature are only moderately lowered as expected from our calculations. Beyond $c/a_{bct} > 1.41$ the samples undergo structural relaxations through adaptive nanotwinning. For all tetragonal structures, we observe a significant increase of the magnetocrystalline anisotropy constant $K_1$, which reaches a maximum of $K_1 \approx -2.4*10^5$ $Jm^{-3}$ at room temperature around $c/a_{bct} = 1.33$ and is thus even larger than for binary $Fe_{70}Pd_{30}$ and the prototype Ni-Mn-Ga magnetic shape memory system. Since $K_1$ represents the driving force for variant reorientation in magnetic shape memory systems, we conclude that Fe-Pd-Cu alloys offer a promising route towards microactuators applications with significantly improved work output.

## Contents

**1. Introduction**

Due to huge strains up to 10% [1] magnetic shape memory (MSM) alloys are of particular interest for microactuators [2]. Most research on bulk and thin films focuses on the Ni-Mn-Ga prototype system [3], but in particular for microsystems $Fe_{70}Pd_{30}$ as the second system discovered [4] shows several advantages. While in the Ni-Mn-Ga system oxidation can result in functional degradation [5] the high content of the noble element Pd in Fe-Pd makes this MSM alloy even biocompatible [6]. Spontaneous magnetic polarisation $J_S$ and Curie temperature $T_C$ are considerably higher [7,8] compared to Ni-Mn-Ga. Whereas the high material cost hinder bulk applications of Fe-Pd, microsystems require only small masses of active material, which makes the material costs negligible compared to the increased process costs of film preparation.

In this paper we quantify the influence of Cu alloying on the magnetic properties of $Fe_{70}Pd_{30-x}Cu_x$. Recent combinatorial experiments revealed that Cu can substantially increase the martensitic transformation temperature [9] which marks the upper limit of the working range for MSM applications. After a thorough crystallographic characterisation of the films we will focus on magnetic aspects, in particular on the magnetocrystalline anisotropy energy. As an intrinsic material property, the anisotropy constant $K$ represents the maximum energy density which can be supplied by an external magnetic field [10]. In case of highly mobile twin boundaries this energy input limits the mechanical work which can be provided by an MSM element. $K$ as the key intrinsic property depends on structure, composition, and chemical order. In addition the following intrinsic magnetic properties have to be considered for MSM applications. A high $J_S$ is favourable, since it allows using low magnetic fields to obtain the maximum energy input at the anisotropy field $H_A = 2K/J_S$. Finally, the Curie temperature should be well above application temperature to allow the MSM effect within an increased temperature regime. Thus, the influence of Cu on spontaneous magnetic polarisation $J_S$ and Curie temperature $T_C$ is investigated using both, experimental as well as theoretical approaches.

To comprehensively characterise materials regarding the above mentioned properties, a high sample quality is required. Single crystals are suitable since finite size, texture and stress effects do not affect the measurement results, but they are difficult to produce and often expensive as large amounts of the material is needed. Epitaxial films represent the thin film counterpart of single crystals and allow to measure the anisotropic magnetic properties directly by measuring the magnetisation curves along different crystallographic directions. This allows easily to determine the magnetocrystalline anisotropy constant $K$ since the films are attached to a rigid substrate that blocks the re-orientation of structural variants. Furthermore, coherent epitaxial growth on cubic seed buffers with different lattice parameters can be used to adjust the tetragonal distortion of the martensite. This approach allows controlling the $c/a_{bct}$ - ratio along the Bain transformation path between almost body centred cubic *bcc* ($c/a_{bct} = 1$) and face centred cubic *fcc* ($c/a_{bct} = 1.41$) [11]. By this kind of epitaxial film growth artificial single variant states are realised [12] allowing to measure $K$ along the different crystallographic directions ($K_1$ and $K_3$).

Recently the authors reported on the straining of epitaxial $Fe_{70}Pd_{30}$ films beyond the Bain path to $c/a_{bct} >1.41$ [13]. For this type of growth, adaptive nanotwinning was found as the relaxation mechanism. By using a simple geometric relation one can still control $K$ by using appropriate buffer lattice constants, but additionally introduces twin boundaries at the nanoscale which can be used as pinning centres e. g. for percolated magnetic recording media. This type of application also requires a high $K$ and benefits from high $J_S$ and $T_C$. In contrast to MSM microactuators, which require a release from the substrate [12] this is not necessary for recording applications.

This paper is arranged as follows: After a short description of sample preparation and methods used for analysis we characterise in detail the structure of the epitaxial films. Supported by extensive first-principles calculations of the elastic energy associated with straining of Fe-Pd and Fe-Pd-Cu films we develop an explanation for the significantly improved growth behaviour of Fe-Pd-Cu. As a second focus we probe the influence of Cu on extrinsic magnetic properties - for example saturation field $\mu_0 H_S$ - as well as the intrinsic magnetic properties like spontaneous polarisation $J_S$ and Curie temperature $T_C$. Finally, rounding up the beneficial impact of Cu on the functional properties of Fe-Pd alloys, we report a significant enhancement of the important magnetocrystalline anisotropy constants $K_1$ and $K_3$ upon addition of Cu.

## 2. Experimental and theoretical methods

Epitaxial thin films were deposited at room temperature using DC magnetron sputtering in a UHV chamber (base pressure <$10^{-8}$ mbar, Argon 6N purity). The Cu content was varied by co-sputtering from a 4 inch $Fe_{70}Pd_{30}$ alloy target (purity 99.99 %,) and a 2 inch Cu target (purity 99.99 %,) and changing the sputter power of Cu (55…60W) and $Fe_{70}Pd_{30}$ (60…55W) targets. With a sputter rate of ~ 0.3 nm s$^{-1}$, we prepared 40 nm thick $Fe_{70}Pd_{30-x}Cu_x$ films. Epitaxial metallic layers (Cr, Au, Ir, Rh, and Cu) used as buffers for the strained growth of $Fe_{70}Pd_{30-x}Cu_x$ films were deposited at 300°C (and at room temperature for Cu) onto MgO(100) single crystal substrates. The nominal film architecture is: MgO/ 50 nm buffer/ 40 nm $Fe_{70}Pd_{30-x}Cu_x$. When metals with *fcc* crystal structure are used as buffer (Au, Ir, Rh, Cu), prior to the deposition, a layer of 5 nm Cr was grown on MgO to enhance adhesion.

X-ray diffraction (XRD, Philips X'Pert) in Bragg-Brentano geometry was performed at room temperature in a two circle setup using Co K$\alpha$ radiation. Pole figures were measured in a four circle setup with Cu K$\alpha$ radiation. The composition of all samples was determined by using a Jeol JSM 5800 LV scanning electron microscope (20 kV acceleration voltage) equipped with an Oxford Inca energy dispersive X-ray (EDX) analysis system (132 eV resolution, every spectrum contained more than 200.000 counts) and a $Fe_{70}Pd_{30}$ standard. EDX was used to probe if the actual composition meets the intended film composition, which had been obtained with a standard deviation of 1 %. In order to determine the thin film thickness, the *Thin Film ID* software from Oxford Instruments was used. This software is based on the *StrataGem* algorithm that allows calculating film thickness from EDX measurements combined with Monte Carlo simulations. After simulation of the electron trajectories within the layers the optimum acceleration voltage is determined and an EDX spectrum is acquired. Further Monte Carlo simulations were separately performed by using Casino software to calculate film thickness [14].

To determine magnetic properties, temperature and magnetic field dependent measurements were performed by using a Physical Property Measurement System (PPMS, Quantum Design) with vibrating sample magnetometer (VSM) add-on. The Curie temperature $T_C$ was determined using Kuz'min's fit [15,16] from temperature-dependent in-plane magnetisation measurements $J_S(T)$ between 50 K and 400 K (sweep rate of 2 K min$^{-1}$) in an applied magnetic field of 1 T:

$$\frac{J_S(T)}{J_0(T=0K)} = \left[1 - s\left(\frac{T}{T_C}\right)^{3/2} - (1-s)\left(\frac{T}{T_C}\right)^{5/2}\right]^{\beta} \qquad (1)$$

For Fe-Pd as a metallic ferromagnetic material we fixed the critical exponent at $\beta = 1/3$ [15]. The other variables – the Curie temperature $T_C$, the saturation polarisation $J_0$, defined as $J_S$ at 0 K, and the shape parameter $s$ are additional variables within this equation and considered as free parameters. For all samples the values for $J_0 = 1 \pm 0.02$ and $s = 1 \pm 0.2$ do not vary significantly compared to the maximum range of $0 < s < 2.5$. This allows to focus on $T_C$ in dependence of tetragonal deformation and composition. MgO single crystals used as substrates contain superparamagnetic impurities that influence the magnetic measurements at low temperatures, with the consequence that $J_0$ was regarded as a free parameter. While for most samples it was sufficient to exclude the temperature range below 50 K for analysis, two samples ($Fe_{70}Pd_{23}Cu_7$ with $c/a_{bct}$ of 1.07 and 1.09) contained impurities which contributed up to 100 K. This reduces the available temperature range significantly; hence these samples had been excluded from the $T_C$ analysis.

The magnetocrystalline anisotropy constants $K_1$ and $K_3$ were calculated from hysteresis measurements in [100], [110] and [001] direction. Magnetic hysteresis loops were measured in the range from -4.5 T up to +4.5 T at room temperature.

Coercivity $H_C$ was determined from the intersection with the field axis, remanence $J_R$ from the intersection with the polarisation axis. Spontaneous polarisation $J_S$ was determined as the maximum polarisation from the in-plane measurements, where the saturation could be obtained in low fields, and from the out-of-plane measurements in high fields. For all samples both values are identical within an accuracy of 3 %. The saturation field $H_S$ was determined by the minimum of the second derivative of the $J(H)$ curves. All values are averaged from positive and negative fields and the difference is about twice the symbol size (when no error bar is shown). This also holds for the systematic error for the field values given which is limited to about $\pm 2$ mT due to the continuous field sweep. The accuracy for the polarisation values is lower and limited by the accuracy of about 5 % for the determination of sample volume.

First-principles calculations in the framework of density functional theory (DFT) allow the accurate and parameter-free determination of structural and electronic properties on the electronic level. In order to obtain information on the binding energy as a function of tetragonality, we performed a full optimisation of the atomic positions without symmetry constraints using the Vienna *ab-initio* simulation package (VASP) [17]. We use the exchange-correlation functional according to Perdew and Wang [18] in combination with the spin interpolation formula of Vosko, Wilk and Nusair [19].The employed projector augmented wave potentials (PAW) describe explicitly the 3$d$ and 4$s$ valence electrons of Fe and Cu and the 4$d$ and 5$s$ of Pd [20]. The cut-off for the plane wave basis was chosen as 342 eV. The Brillouin zone integration was carried out using a 2×2×2 Monkhorst-Pack $k$-mesh with Methfessel-Paxton Fermi-level smearing with a broadening parameter of 0.1 eV. A disordered arrangement was realized using a 500 atom supercell and a random distribution of the 340 Fe atoms and 160 Pd atoms, or respectively, 135 Pd and 25 Cu atoms. In order to avoid statistical uncertainties which hamper a direct comparison, 25 randomly chosen Pd atoms were exchanged by Cu to model a comparable ternary distribution, while the remaining elements were left untouched. Further computational details can be found in [13,21].

In addition, we investigated composition dependent and finite temperature magnetic properties using the Korringa-Kohn-Rostoker (KKR) approach as implemented in the Munich SPR-KKR code (version 5.4) [22, 23]. The exchange-correlation functional was treated within the generalised gradient approximation (GGA) according to Perdew, Burke and Ernzerhof [24] based on a scalar relativistic description of the Hamiltonian in combination with the atomic sphere approximation (ASA). We used 64 points for the energy integration contour in the complex plane while taking into account $f$ states for the angular momentum expansion. We used a dense k-mesh of up to 40×40×40 in the full Brillouin zone, corresponding to 4500 points in the irreducible zone. Here, the disordered nature of the Fe-Pd alloy was modelled in terms of averaging the electronic scattering properties within the coherent potential approximation (CPA). The CPA enables the economic use of small cells but does not allow for structural relaxations. We determined the magnetic exchange parameters for use within a Heisenberg model following the approach by Liechtenstein *et al.* [25], starting from the ferromagnetic state. The corresponding Curie temperatures were estimated within the mean-field approximation.

### 3. Structure and epitaxial relationship

There are several crystallographic structures reported in quenched Fe$_{70}$Pd$_{30}$ bulk. These different structures that appear around room temperature can be described by a phase sequence that changes with increasing Pd content as follows: body-centred-cubic (*bcc*), body-centred-tetragonal (*bct*), face-centred-tetragonal (*fct*) and face-centred-cubic (*fcc*) [7,4,26,27]. The transformation of a crystal lattice from a *bcc* to a *fcc* structure can be described in terms of a tetragonal distortion, following the so-called Bain path [28]. In order to describe the degree of this tetragonal lattice distortion, we will use the $c$ and $a_{bct}$ lattice parameters corresponding to the *bcc* elemental cell. Thus, the ratio $c/a_{bct}$ ranges from 1 (for a perfect *bcc* lattice) over 1.33 (corresponding to a *fct* structure) to 1.41 (for a *fcc* structure). Frequently, the degree of tetragonal distortion is expressed using the austenitic *fcc* elemental cell as basis. Then, the $c/a_{fcc}$ ratio ranges from 1 for the austenite (*fcc* lattice) over 0.94 for the *fct* martensite to 0.71 for *bcc* lattice. The *fcc* description is easily converted to the *bcc* description by multiplying with a conversion factor of √2. Within this paper the $c/a_{fcc}$ ratio of the *bcc* elemental cell is used to describe the lattice deformations, since bulk experiments and DFT calculations indicate that the *bcc* phase defines the ground state [13,27].

The $c/a_{bct}$ ratio in epitaxial films is determined by dividing the values of the out-of-plane *c*-axis by the in-plane *a*-axis of the *bct* cell. Within this chapter the route to determine the *c/a* ratio is described on the basis of the *bcc* unit cell. The $c/a_{bct}$ ratio is used from the beginning to identify the samples and to distinguish between the different tetragonal distortions.

By coherent epitaxial growth the lattice constants of the buffer adjust the in-plane lattice parameters of Fe-Pd-X films and thus $c/a_{bct}$. For analysing the crystal structure of $Fe_{70}Pd_{27}Cu_3$ films grown on different buffer layers, XRD scans in Bragg Brentano geometry were performed (Figure 1). However, this method is only suitable to reveal the out-of-plane lattice parameter. When changing the buffer materials from Au to Cu, the position of the (002) reflection of *fcc* buffers (red circles) increases towards higher 2θ values. By using a *bcc* buffer like Cr the peak position of the (002) reflection is at considerable higher 2θ values. This reflects the shifting of the XRD peak in dependence of the lattice parameter for several buffers layers with different structures. When varying the buffer material, the position of the (002) reflection of $Fe_{70}Pd_{27}Cu_3$ (blue pentagons) shifts from 2θ = 70.5° ($c/a_{bct}$ = 1.07) to a lower value at 2θ = 60.6° ($c/a_{bct}$ = 1.31). For a film on the Cu buffer ($c/a_{bct}$ = 1.57) the (002) reflection shifts to 2θ = 52.6°. From these XRD measurements only the (002) diffraction peak is observed for the $Fe_{70}Pd_{27}Cu_3$ films, which indicates the presence of a highly textured material. The absence of any further diffraction peaks suggests a coherent epitaxial growth of the films, where the constraint of an almost constant unit cell volume results in an increase of the out-of-plane lattice parameter when the in-plane lattice parameters shrink.

To probe this in detail we measured the in-plane lattice parameters by performing 2θ scans of the $(101)_{bct}$ lattice planes in tilted conditions at a tilt angle Ψ. As sketched in Figure 2b a simple geometric relation allows to use this lattice plane (together with the measurements in Bragg-Brentano geometry) to calculate the in-plane lattice parameter. It was found that the in-plane lattice parameters of the $Fe_{70}Pd_{27}Cu_3$ films are identical to the buffer lattice spacings $d_{buffer}$ (Figure 3a). This indicates a coherent growth in all thin films deposited on the different buffers and reveals the following relation: $d_{buffer} = a_{bct}$. The $c/a_{bct}$ ratio for every thin film sample was then calculated by using Bragg-Brentano as well as tilted XRD measurements. This is concluded in Figure 3b, depicting the dependency of the tetragonal deformation ($c/a_{bct}$ ratio) for all buffer materials used. This deformation behaviour is quite similar to previous investigations for thin binary $Fe_{70}Pd_{30}$ films which are added for comparison [11,13]. In contrast to binary Fe-Pd films the $c/a_{bct}$ ratio of $Fe_{70}Pd_{27}Cu_3$ is found to be slightly smaller, suggesting that the addition of smaller atoms like Cu reduce the volume of the $Fe_{70}Pd_{30}$ unit cell.

Thin film samples with varying $c/a_{bct}$ ratio enable us to investigate the change of intrinsic properties as function of the tetragonal deformation and therefore at different structures. In the following chapters we use the variation of the $c/a_{bct}$ ratio in addition to the composition as a control parameter to adjust the magnetic properties. For a detailed understanding of the anisotropic magnetic properties, however, it is first necessary to know how the unit cell is oriented on the buffer.

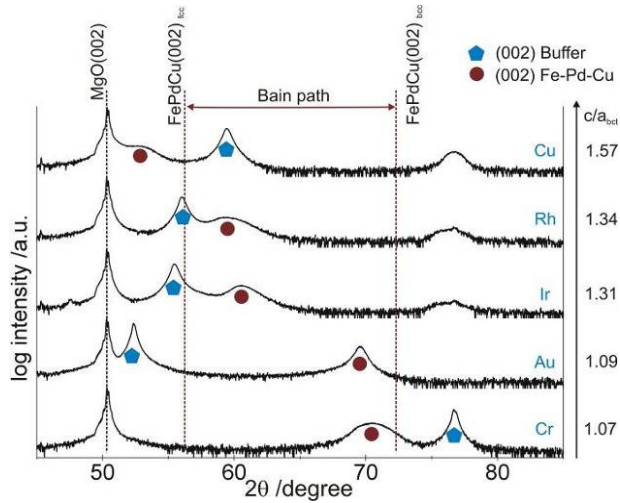

Figure 1: Bragg-Brentano scans of $Fe_{70}Pd_{27}Cu_3$ films on different buffer layers. Red circles mark (002) reflections of the Fe-Pd-Cu films and blue pentagons (002) reflections of the cubic buffers. With increasing $c/a_{bct}$ ratio a shift of $(002)_{Fe-Pd-Cu}$ reflections to lower angles is observed. Red dotted lines indicate the boundaries of the Bain transformation path.

To determine the structural quality of the thin film samples $(101)_{bct}$ pole figure measurements were performed. Due to the four-fold surface symmetry of the (001) oriented MgO substrate only one quadrant of the pole figure measurement for $Fe_{70}Pd_{27}Cu_3$ films on different buffers are depicted in Figure 2a. Within this drawing the MgO [100] edges are oriented parallel to the edges of the figure. This reveals an orientation relationship of the *bct* unit cell of the film and MgO(001) substrate as follows:

$Fe_{70}Pd_{30-x}Cu_x(001)[110] \| fcc\text{-buffer}(001)[100] \| Cr(001)[110] \| MgO(001)[100]$ or

$Fe_{70}Pd_{30-x}Cu_x(001)[110] \| bcc\text{-buffer}(001)[110] \| MgO(001)[100]$.

The corresponding pole figures (Figure 2, illustrate with [29]) indicate a high quality epitaxial growth of the films since well-defined and sharp peaks were observed. The unit cells are rotated by 45° with respect to each other - shift in $\Phi$ direction of about 45°- to reduce misfits between different layer materials.

These pole figure measurements can be also used to determine the $c/a_{bct}$ ratio, since the tilt angle $\Psi$ of the $(101)_{bct}$ plane is connected with the tetragonal distortion by $\tan\psi = c/a_{bct}$ (Figure 2b). This relation predicts a tilt angle of the (101) pole at $\Psi \approx 45°$ for a *bcc* structure ($c/a_{bct} = 1$) and at $\Psi \approx 54°$ for a *fcc* structure ($c/a_{bct} = 1.41$).

When the buffer material is varied, the $\Psi$ shift of the poles reflects the structural changes from *bcc* towards *fcc*. For a film on Cr buffer the (101) pole is at a tilt angle $\Psi \approx 47°$. This confirms an almost cubic structure with $c/a_{bct} = 1.07$. On the other side an Ir buffer layer shifts $\Psi$ to 52°, that gives a $c/a_{bct}$ ratio of 1.31 corresponding to a tetragonal *fct* structure. The pole-figure measurement therefore is a second independent measurement confirming the $c/a_{bct}$ ratios obtained from $\theta-2\theta$ scans with an accuracy of 1 %.

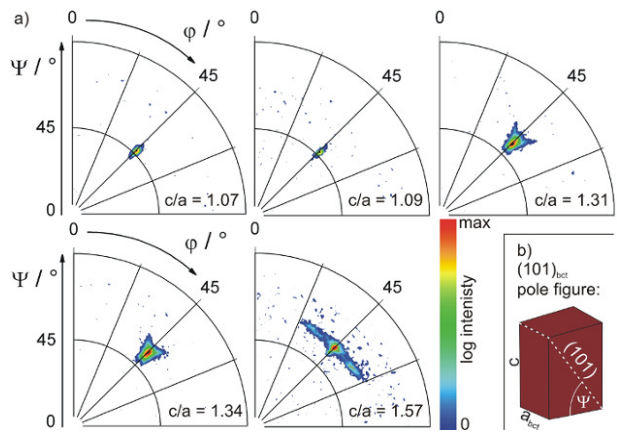

Figure 2: a) $(101)_{bct}$ pole figures of $Fe_{70}Pd_{27}Cu_3$ films on different buffer layers. With increasing $c/a_{bct}$ ratio a shift of the (101) pole towards higher $\Psi$ angles is observed. b) Sketch illustrating the geometric relation between the tetragonal distortion ($c/a_{bct}$ ratio) and the tilt angle $\Psi$ within a *bct* unit cell.

A further increased tetragonal distortion is visible for a film on a Cu buffer ($c/a_{bct} = 1.57$). In Bragg-Brentano scans (Figure 1) the (002) reflection of the $Fe_{70}Pd_{27}Cu_3$ film is observed beyond the limits of the Bain path. This tetragonal distortion agrees well with $c/a_{bct} = 1.57$ derived from the position of the pole in the $(101)_{bct}$ pole figure at $\Psi \approx 57°$. The intensity in this measurement further exhibits a substantial broadening and splitting in Φ direction. This observation agrees well with the authors' previous report on highly-strained binary $Fe_{70}Pd_{30}$ films on Cu buffer layers [13]. The broadening can be ascribed to adaptive nanotwinning that reduces the huge elastic energy induced by the coherent epitaxial strain through the formation of twin boundaries. The rotation of the tetragonal variants which form $(101)_{fct}$ twin boundaries leads to the small observed additional intensities.

The same set of XRD measurements was performed for a series of films with increased Cu content. These measurements are not shown here, since they are practically identical to the ones obtained for the lower Cu content. From these measurements we conclude that even at a Cu content of 7 at.% no macro-precipitations occur in epitaxial films. This is different to polycrystalline annealed Fe-Pd-Cu thin films which decompose at Cu contents above 4 at.% [9]. We attribute this to the difference in fabrication process routes. Epitaxial films are deposited at room temperature without any post annealing which avoids macro-precipitates compared to annealed films.

The lattice parameters determined for $Fe_{70}Pd_{23}Cu_7$ are equal to those of $Fe_{70}Pd_{27}Cu_3$. No change in tetragonal deformation with increased Cu content is observed. Accordingly we do not distinguish between 3 and 7 at.% Cu in Figure 3. Experiments on polycrystalline annealed films have shown that with increasing Cu content the lattice constant decreases [9], since the atomic radius for Cu (0.128 nm) is smaller than for Pd (0.137 nm). However, it is known that alloys, like Fe-Pd, exhibiting the Invar effect [30,31,32], deviate from the rule of mixture of Vegard's law [33,34]. This must be considered for the Fe-Pd and the second quasi binary Cu-Pd systems [35].

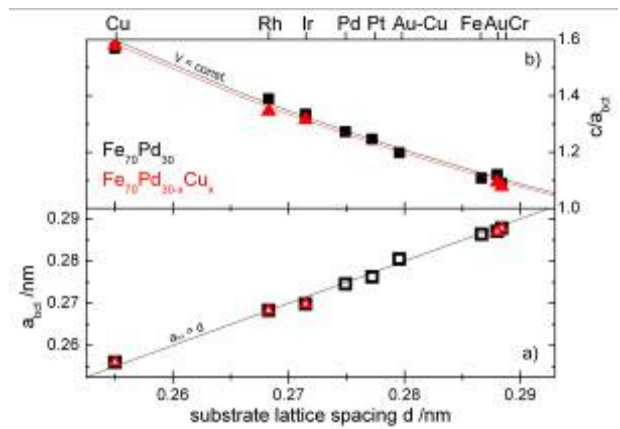

Figure 3: a) In-plane lattice parameters $a_{bct}$ for $Fe_{70}Pd_{30}$ (rectangle) and $Fe_{70}Pd_{30-x}Cu_x$ (triangle) are fixed by the substrate's lattice spacing $d$ of different buffers (as marked on the top). b) The change of tetragonal deformation ($c/a_{bct}$) dependent on buffer material is smaller for the $Fe_{70}Pd_{30-x}Cu_x$ system (triangle) than for the binary $Fe_{70}Pd_{30}$ (rectangle [11,13]). The curves illustrate the expected change in deformation at constant volume of the unit cell.

The improved growth behaviour can be motivated in terms of the structural variation of the total energy. We obtain this energy from first-principles within a supercell description involving a full relaxation of the atomic position which minimises the interatomic forces for each tetragonal stage. According to Figure 4a, only a small variation of elastic energy $E$ is involved if we vary the $c/a_{bct}$ ratio along the entire Bain transformation path. In addition to the global minimum at *bcc* we observed a second local minimum beyond the Bain path ($c/a_{bct} > 1.41$). Such a local minimum can be attributed to the formation of a nanotwinned microstructure in the simulation cell [13] and is further confirmed by our pole figure measurements depicted previously in this chapter (Figure 2). This minimum is observed for the binary $Fe_{68}Pd_{32}$ solid solution, but also for ternary $Fe_{68}Pd_{27}Cu_5$, where the feature is comparatively shallow. Replacing 16% of Pd atoms by Cu increases the overall valence electron number $e/a$ by 0.05, which by trend stabilises the *fcc* austenite [36,31,32]. This is reflected in our calculations by the noticeably decreased energy difference between the *bcc* ground state at $c/a_{bct} = 1$ and the *fcc* austenite at $c/a_{bct} = 1.41$. Interestingly, the energy profile does not change significantly at low energies for $c/a_{bct} < 1.3$. Changes appear mainly in the vicinity of the local *fcc* maximum, which is now embedded in flatter energy landscape. A flat profile in the vicinity of the martensitic phase, however, can be considered beneficial for the formation of the metastable *fct* phase, which is required for the MSM behaviour. The decrease of the *fcc-bcc* energy difference implies a decreased temperature for the onset of the martensitic transformation, but this can be compensated by the simultaneous replacement of Pd by Fe, as shown in a previous study [9]. The simultaneous substitution of Pd by Cu and Fe therefore opens a way to specifically design the profile of the binding surface around $c/a_{bct} = \sqrt{2}$ and thus the stability range of the *fct* phase.

The total difference in elastic energy between *bcc* and *fcc* state $\Delta E$ is reduced from $\Delta E = 15$ meV/atom to $\Delta E = 12$ meV/atom for $Fe_{68}Pd_{32}$ and $Fe_{68}Pd_{27}Cu_5$, respectively. This reduces the driving force for the transformation and all associated relaxation processes and is thus favourable for strained epitaxial growth. Indeed, to avoid unfavourable relaxation mechanism like $(111)_{fcc}$ deformation twinning [37], very low deposition rates of 0.024 nm s$^{-1}$ were required for epitaxial growth of binary $Fe_{70}Pd_{30}$ films [12]. The present Fe-Pd-Cu films, however, were grown with a deposition rate one order of magnitude higher (0.3 nm s$^{-1}$). Still the poles in the (101) pole figure and the diffraction peaks in the Bragg-Brentano measurement reveal a small full-width-at-half-maximum (FWHM). By using a modified Scherrer equation [38] the determined coherence length is in the range of the overall thickness of the Fe-Pd-Cu films, confirming a high quality epitaxial growth. This allows producing thick epitaxial $Fe_{68}Pd_{30-x}Cu_x$ films, as required for microsystems, in a shorter time.

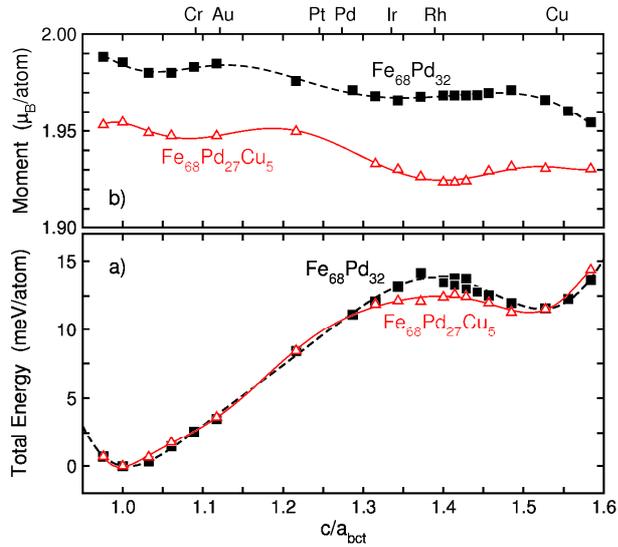

Figure 4: a) Variation of the total energy per atom as a function of the tetragonal distortion $c/a_{bct}$ obtained from *ab-initio* calculations of a $Fe_{68}Pd_{32}$ (black squares) and $Fe_{68}Pd_{27}Cu_5$ supercell (red triangles). The calculations were carried out at constant volume of 13.1 Å$^3$/atom and include a full geometric optimisation of the atomic positions. Apart from the *bcc* ground state minimum at $c/a_{bct} = 1$ (which defines the energy reference for each composition), in both cases a second local minimum around $c/a_{bct} = 1.5$ (beyond *fcc*) is obtained which corresponds to the appearance of a finely twinned adaptive superstructure in the 500 atom simulation cell. b) The variation of the average ground state magnetic moment (normalised per atom) for both configurations as a function of $c/a_{bct}$.

## 4. Remanence, coercivity and saturation field

The hard magnetic axis in a tetragonal lattice of a bulk $Fe_{70}Pd_{30}$ single crystal is aligned along the *c*-axis while the magnetic easy directions lie within the basal plane [7,39]. Within the easy plane, not all directions are equivalent and a slight anisotropy is observed, which favours both $[110]_{bct}$ easy axes.

To investigate structure-dependent magnetic properties of epitaxial $Fe_{70}Pd_{27}Cu_3$ films, in-plane hysteresis curves for thin film samples with different tetragonal deformation were measured at room temperature (Figure 5a-e). The magnetisation behaviour along the different crystallographic in-plane directions - $[100]_{bct}$ and $[110]_{bct}$ – differ significantly as known also for binary $Fe_{70}Pd_{30}$ films. The differences are most evident for $c/a_{bct} = 1.31$, which is close to the middle of the Bain transformation path. Here, both crystallographic directions show different values for remanent polarisations $J_R$ and saturation fields $H_S$, while the coercivity fields $H_C$ do not vary significantly.

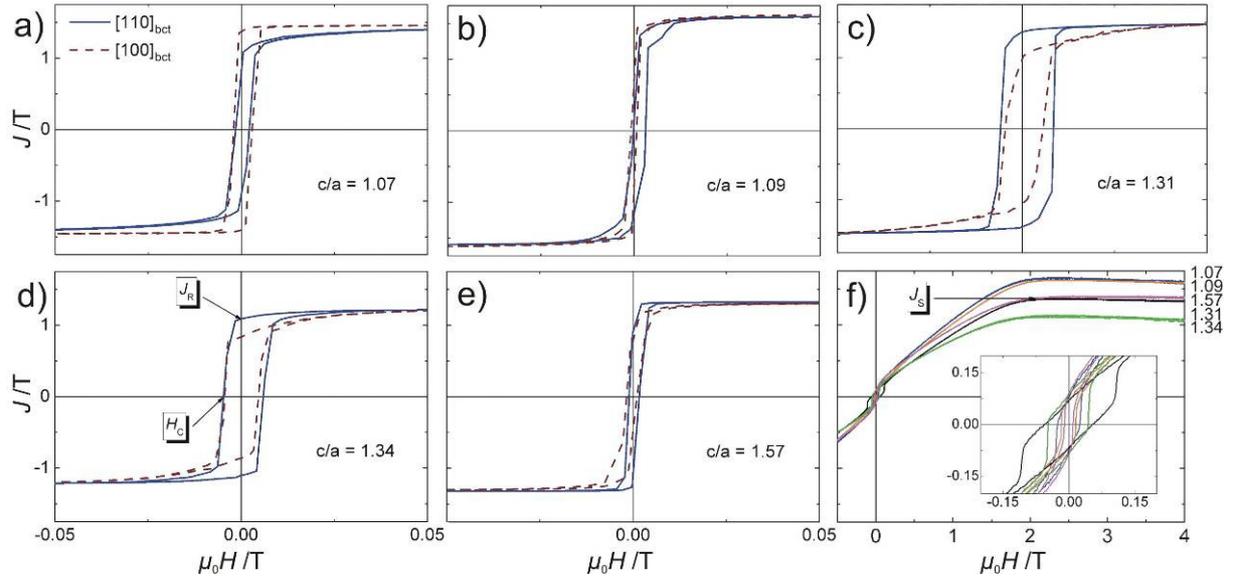

Figure 5: Probing magnetocrystalline anisotropy within the basal plane by measuring hysteresis curves for different tetragonal deformation ratios measured at 300 K (a - e). Solid lines show the measurements along $[110]_{bct}$ direction and broken lines along $[100]_{bct}$ direction. f) Comparison of hysteresis loops measured out-of-plane along $[001]_{bct}$ at 300 K. The inset shows a zoom around $\mu_0 H = 0$ T. From these measurements, the magnetic constants $J_R$, $J_S$ and $H_C$ were determined.

In dependence of the $c/a_{bct}$ ratio the shape of magnetisation curves and thus the magnetic characteristics change. At low tetragonal deformation ($c/a_{bct} = 1.07$) the hysteresis measured in $[100]_{bct}$ direction shows a step like switching behaviour. In the $[110]_{bct}$ direction a similar switching process occurs, but additionally a nearly linear increased magnetisation at higher fields is observed. This behaviour originates from magnetisation rotation, indicating that the $[110]_{bct}$ direction is the harder axis within the basal plane. Accordingly, $J_R$ is reduced in this direction. When increasing the $c/a_{bct}$ ratio to 1.31 and above, the opposite behaviour is observed. This can be correlated to a change in sign of the magnetic anisotropy within the basal plane. The extracted values for the coercivity field $H_C$ and saturation field $H_S$ along $[110]_{bct}$ direction are summarised in Figure 6a and b.

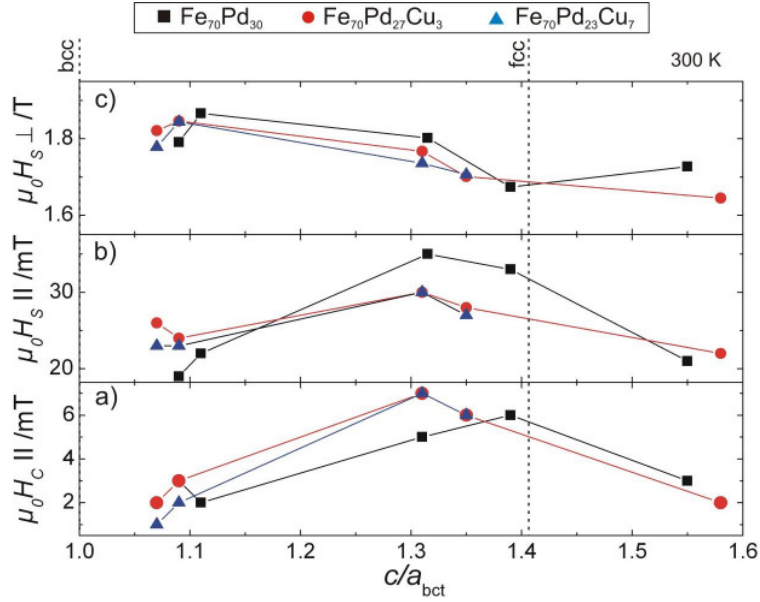

Figure 6: $c/a_{bct}$ dependency of in-plane coercivity field $\mu_0 H_C \parallel [100]_{bct}$, (a) and saturation field $\mu_0 H_S \parallel [100]_{bct}$ (b) as well as the out-of-plane saturation field $\mu_0 H_S \parallel [001]_{bct}$ (c) extracted from the hysteresis curves measurements in Figure 5. Black rectangles represent the results for $Fe_{70}Pd_{30}$ films, red circles $Fe_{70}Pd_{27}Cu_3$ films and blue triangles $Fe_{70}Pd_{23}Cu_7$ films. All lines are guides for the eye.

The trend of $\mu_0 H_C$ and $\mu_0 H_S$ within the film plane for different tetragonal distortions is similar to epitaxial $Fe_{70}Pd_{30}$ films (black symbols in Figure 6) and previous work on epitaxial $Fe_{1-x}Pd_x$ films deposited on MgO(100) substrates [40,41]. Due to high crystal symmetry, films close to the *bcc* structure exhibit a low $H_C$ and $H_S$ while films with a *bct* structure have a slightly increased $H_C$ and slightly reduced $H_S$. *Fct* films exhibit a different loop shape with highest $H_C$ and $H_S$ due to the large lattice deformation. Close to the *fcc* phase $H_C$ and $H_S$ are reduced because of the high crystal symmetry of the *fcc* structure. $Fe_{70}Pd_{27}Cu_3$ films also follow this behaviour. Both values $H_C$ and $H_S$ approach zero when the tetragonal distortions is close to the cubic structures *bcc* ($c/a_{bct} = 1$) and *fcc* ($c/a_{bct} = 1.41$). For the tetragonal structures both values show a maximum.

For a huge tetragonal deformation of $Fe_{70}Pd_{30}$ and $Fe_{70}Pd_{30-x}Cu_x$ beyond the Bain path ($c/a_{bct} = 1.57$, Figure 5e), $H_C$ decreases again and only a slight differences between both in-plane curves are observed. This behaviour will be discussed together with the magnetic anisotropy constants $K_1$ and $K_3$ in chapter 6.

To obtain the magnetic anisotropy constant $K_1$, out of plane hysteresis curves along $[001]_{bct}$ direction were measured (Figure 5f). Along in-plane directions, all films are fully saturated at a magnetic field of 0.1 T, while for out-of-plane measurements substantially higher fields above 1 T are required. Shape anisotropy dominates the magnetisation behaviour for an ideal thin film (demagnetisation factor $N = 1$). When neglecting magnetocrystalline anisotropy the magnetic saturation is expected at $\mu_0 H_S = J_S$. The difference between $\mu_0 H_S$ and $J_S$ will be later used to calculate the magnetic anisotropy constant $K_1$. When increasing the tetragonal deformation to $c/a_{bct} \leq 1.41$, $H_S$ decreases (Figure 6a).

For all measurements a small hysteresis is observed (inset in Figure 5f). We attribute this switching process to a slightly angular misalignment of the sample normal with respect to the field. For a pinning controlled coercivity mechanism the coercivity exhibits a Kondorsky like strong increase when the field direction approaches hard direction [42]. At the same time the remanence to saturation ratio decreases. As a detailed analysis requires a more accurate control of the tilt angle, we do not give $H_C$ from the out-of-plane direction here.

We extracted values for $H_C$ and $H_S$ along $[110]_{bct}$ and $[001]_{bct}$ direction also from the in-plane and out-of-plane hysteresis curves for $Fe_{70}Pd_{23}Cu_7$ films, which follow the same trend as films with 3 at.% Cu (see blue triangles in Figure 6).

## 5. Change of Curie temperature and spontaneous polarisation

Previous experiments have shown that a tetragonal distortion of the lattice also significantly affects the Curie temperature $T_C$ [11]. Temperature dependent magnetisation curves were measured in-plane along the $[110]_{bct}$ direction. In this direction an applied field of 1 T is sufficient to saturate the sample. Unfortunately, the magnetisation curves can only measured up to 400 K, which is below $T_C$, to avoid the destructive thermal decomposition of the metastable alloy [43]. Therefore, we were forced to determine $T_C$ of our films by an extrapolation of the magnetisation curves using Kuz'min's empirical fit [15,16] as explained in detail in Section 2. The results are shown in Figure 7a.

Black rectangles represent the values for $Fe_{70}Pd_{30}$, red circles $Fe_{70}Pd_{27}Cu_3$ and blue triangles $Fe_{70}Pd_{23}Cu_7$. The $Fe_{70}Pd_{30}$ as well as the $Fe_{70}Pd_{30-x}Cu_x$ systems show a similar behaviour. Increasing $c/a_{bct}$ ratio up to 1.41 $T_C$ decreases monotonously. The value of $T_C = 652$ K at $c/a_{bct} = 1.39$ approaches the literature value of 600 K reported for the bulk *fcc* phase ($c/a_{bct} = 1.41$) [8,44]. The change of $T_C$ by addition of Cu is within the experimental error range. Thus, compared to other MSM systems like Ni-Mn-Ga ($T_C < 370$ K), $T_C$ is still much higher [45].

An independent access to the compositional and structural trends concerning $T_C$ and ground state $J_S$ is provided by our *ab-initio* calculations which corroborate the above experiments. $T_C$ is obtained from a classical Heisenberg model which is parameterised with first principles magnetic exchange constants. In the present case, the mean-field approximation can already give a reasonable estimate of the structural and compositional trends governing the magnetic transformation. Nevertheless, one must be aware that this simple approach systematically overestimates critical temperatures (typically by 20...30%) due to the neglect of spin-fluctuations. Furthermore, also the induced nature of the Pd moments, which vary in magnitude according to the field of the surrounding localised Fe moments, is not taken into account, either [46]. But a comparison with a numerically exact Monte-Carlo treatment of the Heisenberg model [47] demonstrates that for a heuristic prediction these shortcomings can be compensated sufficiently by correcting $T_C$ systematically with an empirical factor of 0.75. The almost quantitative agreement which we obtain along the Bain path between *bcc* and *fcc* (Figure 7b) nicely corroborates the experimental procedure, whereas we find no significant variation of $T_C$ with the Cu-content.

For $c/a_{bct} > 1.41$ - beyond the Bain path - the experimental $T_C$ increases again notably. A $T_C$ of 822 K at $c/a_{bct} = 1.56$ is similar to values for a structure with $c/a_{bct} = 1.11$. In contrast, our calculations assuming a single variant with $c/a_{bct} = c/a_{box} > 1.41$ according to the substrate constraint exhibit only a small increase in $T_C$ (solid line in Figure 7b). The agreement is much better, if we assume an energetically more favourable nanotwinned *fct* configuration, where the individual twins are rather described by configurations with $c/a_{bct} < 1.41$. In a recent work, we derived a simple approximate relation between the epitaxial constraint from the substrate $c/a_{box}$ resulting and the tetragonality of the twins $c/a_{twin}$ [13]:

$$\left.\frac{c}{a}\right|_{box} = \sqrt{1 + 2\left(\left.\frac{c}{a}\right|_{twin}\right)^{-2}}. \qquad (2)$$

Using this relation to extrapolate our values for $c/a_{bct} > 1.41$ beyond *fcc*, we obtain a reasonable agreement with experiment concerning the Cu buffer (dashed line in Figure 7b).

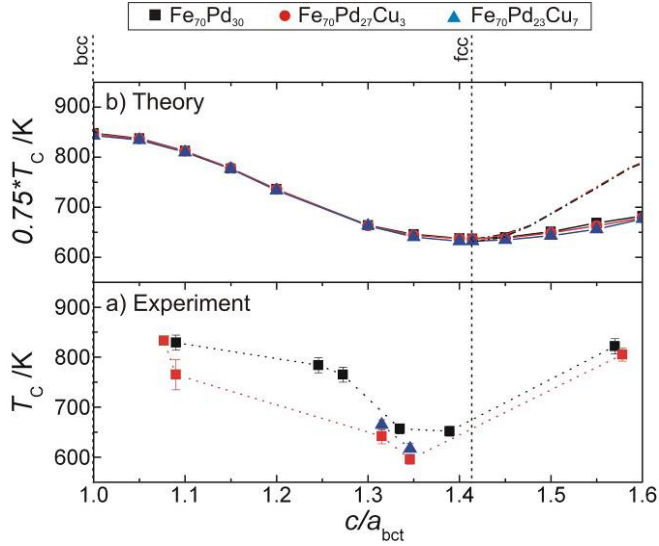

Figure 7: Experimental (a) and calculated (b) Curie temperature $T_C$ in dependence of $c/a_{bct}$ and composition. With increasing $c/a_{bct} < 1.41$ $T_C$ decreases. For $c/a_{bct} = 1.57$ similar values as at $c/a_{bct} = 1.11$ are observed. The theoretical Curie temperatures are obtained using the mean-field approximation to the Heisenberg model and consistently corrected by a factor 0.75. The dashed lines in b) represent results for the adaptive nanotwinning concept. These data agree well with the experimental values.

Another crucial intrinsic parameter for the MSM effect is a high spontaneous magnetic polarisation $J_S$. Figure 8a shows $J_S$ for the different Cu contents in dependence of the tetragonal deformation. We extracted these values from the deformation dependent changes in the hysteresis curves in Figure 5f. Similar to the structural variation of $T_C$, we observe a minimum of $J_S = 1.19$ T at $fcc$ ($c/a_{bct} = 1.41$). Close to $bcc$ ($c/a_{bct} = 1.09$) and at huge strains, $J_S$ reaches values of 1.76 T.

Both, $Fe_{70}Pd_{27}Cu_3$ and $Fe_{70}Pd_{23}Cu_7$ films, reach just 80 % of the $J_S$ of $Fe_{70}Pd_{30}$, which is in agreement with previous experiments on polycrystalline samples [9]. Nevertheless, the values for $J_S$ are still much higher than 0.76 T obtained for the MSM prototype system Ni-Mn-Ga [48].

While the main trends are identical - a decrease of $J_S$ with increasing Cu content and a minimum at $fcc$ - the experimental values obtained at ambient conditions (300 K) close to $T_C$ exhibit a significantly stronger variation with both, composition and $c/a_{bct}$ ratio, than the results from the *ab-initio* calculations at $T = 0$ K (Figure 8b). Here, the ground state magnetic moments vary by a few percent region while the changes in experiment are almost one order of magnitude larger. A larger reduction at $fcc$ is expected at ambient conditions, since $fcc$ exhibit the lowest $T_C$. It may also be speculated whether longitudinal spin fluctuations, which can occur in Invar materials as $fcc$ $Fe_{70}Pd_{30}$ [31,32], reduces $T_C$ in Austenite further.

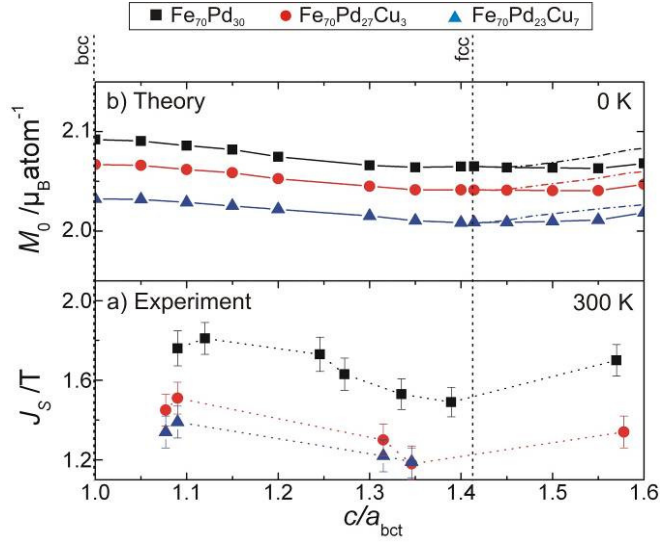

Figure 8: a) Spontaneous polarisation $J_S$ determined for different $c/a_{bct}$ ratios and compositions at 300 K. $J_S$ for $Fe_{70}Pd_{30-x}Cu_x$ is approximately 80 % of the value for $Fe_{70}Pd_{30}$. All lines are guides for the eye. b) Calculated ground state total (spin + orbital) magnetic moments in dependence of $c/a_{bct}$ and composition at 0 K. The dashed lines represent results for the adaptive nanotwinning concept.

## 6. Change of magnetocrystalline anisotropy energy

The key intrinsic magnetic property of MSM alloys is their magnetocrystalline anisotropy energy (*MAE*). During the MSM effect the *MAE* represents the maximum energy input possible by a applying an external magnetic field. Hence the *MAE* limits the energy available to move twin boundaries or conduct external work. *MAE* of a tetragonal lattice can be described by the following equation [49]:

$$MAE = K_1 \sin^2 \alpha + K_2 \sin^4 \alpha + K_3 \sin^4 \alpha \cos 4\beta. \tag{3}$$

$K_i$ are the anisotropy constants, $\alpha$ is the angle between magnetisation direction and *c*-axis and $\beta$ is the angle to the *a*-axes within the basal plane of a tetragonal lattice. In most compounds without rare-earth elements higher order terms ($K_2$) can be neglected. For the $Fe_{70}Pd_{30}$ system this was shown by Cui *et al.* [7]. Hence it is sufficient to consider here only $K_1$ which describes the work required to magnetise the sample along the hard magnetisation *c*-axis and $K_3$ which describes the four-fold anisotropy within the basal plane.

In this paper the $c/a_{bct}$ ratio was controlled by strained epitaxial film growth. This allows to determine $K_i$ for all distortions at one temperature (here: room temperature), which is not possible in bulk. Due to the martensitic transformation and additional strong temperature dependency of the $c/a$ ratio, it is not possible to separate between the influence of *T* and *c/a* in bulk. From hysteresis measurements (Figure 5) we know, that in our epitaxial films the [001]$_{bct}$ direction is the hard magnetisation axis and [100]$_{bct}$ and [110]$_{bct}$ directions form the easy plane.

From the present thin film experiments $K_1$ can be extracted from the measurements of $H_S$ along the hard [001]$_{bct}$ direction. The shape anisotropy was considered by using a demagnetisation factor $N = 1$ of an ideal infinite film and the anisotropy field $H_A$ can be calculated by $\mu_0 H_A = \mu_0 H_S - N J_S$. The anisotropy field is then converted in $K_1 = -\frac{H_A \cdot J_S}{2}$.

In Figure 9a $K_1$ is plotted as a function of the $c/a_{bct}$ ratio and Cu content. The tetragonal deformation of a Fe-Pd unit cell results in the formation of an easy plane, i.e., $K_1 < 0$. The maximum effect is observed around $c/a_{bct} = 1.33$ corresponding to the *fct* structure, which exhibits the MSM effect in bulk. For binary *fct* $Fe_{70}Pd_{30}$ we find $K_1 = -1.6*10^5$ Jm$^{-3}$. This agrees well with literature values reported for *fct* single crystals (open rectangles in Figure 9a) and DFT calculations (stars in Figure 9a) [50,40]. In contrast, $Fe_{70}Pd_{30-x}Cu_x$ films exhibit an increased absolute value of magnetocrystalline anisotropy of $K_1 \sim -2.4*10^5$ Jm$^{-3}$.

The absolute values for $K_1$ of $Fe_{70}Pd_{27}Cu_3$ and $Fe_{70}Pd_{23}Cu_7$ even exceed the magnetocrystalline anisotropy constants reported for the Ni-Mn-Ga system at room temperature: $K_1 = 1.65*10^5$ Jm$^{-3}$ for a 10M single variant single crystal and $K_1 = 1.7*10^5$ Jm$^{-3}$ for the 14M structure [45]. A larger MAE of similar magnitude is only observed for Ni-Mn-Ga at significantly lower temperatures [48, 51] according to the considerable variation of the MAE with temperature in uniaxial magnets [52].

When changing the tetragonal deformation close to highly symmetric cubic structures (*bcc* and *fcc*,) $K_1$ is reduced for all compositions. For $c/a_{bct}$ larger than 1.41, magnetocrystalline anisotropy increases again, but does not reach the values of *fct*. This is again consistent with the presence of adaptive nanotwins in the film [13]. For a very small variant width it is not anymore possible to form a complete 90° domain wall at twin boundaries since exchange energy favours a parallel alignment of magnetisation. According to the random anisotropy model of Herzer the critical length for this is the magnetic exchange length $l_{exch}$ [53]. For *fct* Fe-Pd $l_{exch}$ is in the order of 18-85 nm (depending on compositions) [40], which far exceeds the width of adaptive nanotwins [13].

Due to the fourfold symmetry of the basal plane a deviation from an idealised easy plane behaviour can be observed in films (Figure 5) and bulk samples [7,39]. $K_3$ is the measure of the anisotropy within this basal plane and defines the work which is necessary to magnetise along the respective directions of the *bct* unit cell: $2K_3 = W[100]_{bct} - W[110]_{bct}$. It can be extracted from magnetisation curves (Figure 5) from the area enclosed by the hysteresis curves measured along both, $[110]_{bct}$ and $[100]_{bct}$ directions.

The values of $K_3$ are two orders of magnitude smaller than for $K_1$ (Figure 9b) and change sign within the Bain path. For $c/a_{bct}$ values close to *bcc* positive values are observed. Films with $c/a_{bct}$ ratios close to *fcc* have negative $K_3$ values. In between these values no significant differences were observed for $Fe_{70}Pd_{30}$ and $Fe_{70}Pd_{30-x}Cu_x$ films. As for $|K_1|$, $|K_3|$ has a maximum with $1.8*10^3$ Jm$^{-3}$ at the *fct* structure ($c/a_{bct}$ = 1.33).

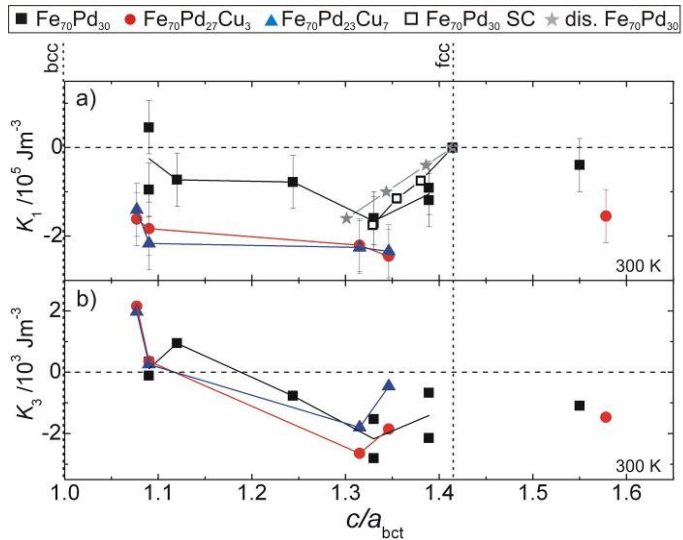

Figure 9: Magnetocrystalline anisotropy constants $K_1$ (a) and $K_3$ (b) as a function of $c/a_{bct}$ for various Cu contents (solid symbols, $T$ = 300 K). Also shown are the values for a bulk $Fe_{68.8}Pd_{31.2}$ single crystal (open rectangle [50]) and calculations for disordered $Fe_{70}Pd_{30}$ (stars, $T$ = 0 K [40]). The errors for $K_1$ are shown by error bars and for $K_3$ they are in the range of the symbol size. All lines are guides for the eye.

## 7. Conclusions

Our detailed analysis of structure and magnetism in ternary Fe-Pd-Cu epitaxial thin films suggests that addition of small amounts of Cu is suitable to significantly enhance the functional properties of the $Fe_{70}Pd_{30}$ magnetic shape memory alloy. We selected Cu since recent experiments on polycrystalline films and splats show that it can increase the maximum solvable Fe-content and thus the martensitic transformation temperature [9]. Our combined experimental and computational approach yields complementary insight into the frozen stages of the martensitic transformation process within the limits of the Bain path, which is enforced by the epitaxial relation to carefully selected buffer materials, This is in particular favourable for a detailed analysis of the anisotropic magnetic properties since the measurements are neither affected by a magnetically induced reorientation nor the continuous variation of the tetragonal distortion with temperature which are both occur in bulk samples.

One central result is that Cu flattens the energy landscape, which suppresses common relaxation mechanisms and thus allows for a much better film quality in combination with faster growth. More important, we observe an increase of the magnetocrystalline anisotropy constant $K_1$ by 40 %, which is a substantial improvement of the key intrinsic property for the MSM effect. The values obtained for the *fct* structure even exceed those reported for the prototype Ni-Mn-Ga system, which makes the Fe-Pd-Cu system of particular interest for microsystems with a high energy density. As a minor drawback, Cu alloying reduces the room temperature spontaneous polarisation by about 15-20 %.

The XRD measurements indicate that the use of room temperature deposition techniques prevents macroscopic demixing, which must be expected for more than 5 at.% Cu, otherwise [9]. However, upon increasing the Cu content from 3 at.% to 7 at.% we do not observe a variation of the magnetic properties. This might in turn be taken as an indication for the presence of structural or compositional inhomogeneities at the nanoscale (e.g., a tendency towards a slight $L1_0$ short range order with antiphase boundaries). Recent Mößbauer experiments on Fe-Pd-Cu splats with < 1 at % Cu suggest some chemical short-range order [54]. Moreover, first principles calculations of Fe-rich Fe-Pd on the other hand predict a certain preference for forming a layered type of order ($L1_0$ or Z1) [55,56], while the cubic $Fe_3Pt$-type $L1_2$ order appears to be ruled out for energetic reasons [21]. In the martensitic state, aging behaviour resulting to symmetry-conforming short-range order at this length scale does not necessarily inhibit shape memory applications [57] and could tentatively induce a kind of beneficial two-way behaviour. This aspect, however, cannot be probed at the films tested with our methods available and requires further investigation.

**Acknowledgments**

The authors thank P. Entel, J. Buschbeck, A. Backen, A. Diestel and R. Niemann for discussions and experimental support and the DFG for funding via the Priority Program SPP 1239. The authors would like to thank the John von Neumann Institute for Computing (NIC), Jülich Supercomputing Center (JSC), both of Forschungszentrum Jülich, as well as the Center for Computational Sciences and Simulation (CCSS) of the University of Duisburg-Essen for computing time and support.

**References**

[1] Sozinov A, Likhachev A A, Lanska N and Ullako K 2002 *App. Phys. Lett.* **80** 1746

[2] Kohl M, Yeduru S R, Khelfaoui F, Krevet B, Backen A, Fähler S, Eichhorn T, Jakob G and Mecklenburg A 2010 *Mat. Sci. Forum* **635** 145

[3] Ullako K, Huang J K, Kantner C, O'Handley R C and Kokorin V V 1996 *Appl. Phys. Lett.* **69** 1966

[4] James R D and Wuttig M 1998 *Philos. Mag.* A **77** 1273

[5] Gebert A, Roth S, Oswald S and Schultz L 2009 *Corros. Sci.* **51** 1163

[6] Ma Y, Zink M and Mayr S G 2010 *Appl. Phys. Lett.* **96** 213703

[7] Cui J, Shield T W and James R D 2004 *Acta Mater.* **52** 35

[8] Kussmann A and Jessen K 1962 *J. Phys. Soc. Jpn.* **17** 136

[9] Hamann S *et al.* 2010 *Acta Materialia* **58** 5949

[10] Heczko O 2005 *J. Magn. Magn. Mater.* **290** 787

[11] Buschbeck J, Opahle I, Richter M, Rößler U K, Klaer P, Kallmaye M, Elmers H J, Jakob G, Schultz L and Fähler S 2009 *Phys. Rev. Lett.* **103** 216101

[12] Bechtold C *et al.* 2010 *Adv. Mat.* **22** 2668

[13] Kauffmann-Weiss S, Gruner M E, Backen A, Schultz L, Entel P and Fähler S 2011 *Phys. Rev.Lett.* **107** 206105

[14] Drouin D, Hovington P and Gauvin R 1997 *Scanning* **19** 20

[15] Kuz'min M D 2005 *Phys. Rev Lett.* **94,** 107204

[16] Kuz'min M D, Richter M and Yaresko A N 2006 *Phys. Rev. B* **73**, 100401

[17] Kresse G and Furthmüller J 1996 *Phys. Rev. B* **54** 11169

[18] Perdew J P 1991 *Electronic Structure of Solids* vol **91**, ed P Ziesche and H Eschrig (Berlin: Akademie)

[19] Vosko S H, Wilk L and Nusair M 1980 *Can. J. Phys.* **58** 1200

[20] Kresse G and Joubert D 1999 *Phys. Rev. B* **59** 1758

[21] Gruner M E and Entel P 2011 *Phys. Rev. B* **83** 214415

[22] The Munich SPR-KKR package, version 5.4, Ebert H *et al.* http://olymp.cup.uni- muenchen.de/ak/ebert/SPRKKR

[23] Ebert H 2000 *Electronic Structure and Physical Properties of Solids (Lecture Notes in Physics* vol 535*)* ed H Dreyss´e (Berlin: Springer) p 191

[24] Perdew J P, Burke K and Ernzerhoff M 1996 *Phys. Rev. Lett.* **77** 3865

[25] Liechstenstein A I, Katsnelson M I, Antropov V P and Gubanov V A 1987 *J. Magn. Magn. Mater* **67** 65

[26] Inoue S, Inoue K, Fujita S and Koterazawa K 2003 *Mater Trans* **44** 298

[27] Sugiyama M, Oshima R and Fujita F E 1984 *Trans. Jpn. Inst. Met.* **25** 585

[28] Bain E C 1924 *T. Am. I. Min. Met. Eng.* **70** 25

[29] Kauffmann A 2011 CorrVert, http://sco.ifw-dresden.de/

[30] Weiss R J 1963 *Proc. Phys. Soc.* **82** 281

[31] Pepperhoff W and Acet M 2001 *Engineering Materials* vol VIII (Berlin: Springer)

[32] Wassermann E F 1990 *Ferromagnetic Materials* vol 5, ed K H J Buschow and E P Wohlfarth (Amsterdam: Elsevier) p 237

[33] Okamoto H 1993 *Phase diagrams of binary iron alloys* (Materials Park Ohio: ASM International)

[34] Kaufman L, Clougherty E V and Weiss R J 1963 *Acta Met.* **11** 323

[35] v. Steinwehr H E 1967 *Zeitschrift für Kristallographie* **125** 360

[36] Nishiyama Z 1978 *Martensitic Transformation* (New York: Academic Press)

[37] Edler T, Buschbeck J, Mickel C, Fähler S and Mayr S G 2008 *New J. Phys.* **10** 063007

[38] Patterson A L 1939 *Phys. Rev.* **56** 978

[39] Kakeshita T and Fukuda T 2006 *Int. J. Appl. Electromagn. Mech.* **23** 45

[40] Buschbeck J, Opahle I, Fähler S, Schultz L and Richter M 2008 *Phys. Rev.* B **77** 174421

[41] Buschbeck J, Hamann S, Ludwig A, Holzapfel B, Schultz L and Fähler S 2010 *J. Appl Phys* **107** 113919

[42] Kondorsky E 1940 *J. Phys.* (USSR) **2**, 161

[43] Buschbeck J, Heczko O, Ludwig A, Fähler S, Schultz L 2008 *J. Appl. Phys.* **103**, 07B334.

[44] Matsui M, Shimizu T, Yamada H and Adachi K 1980 *J. Magn. Magn. Mater.* **15** 1201

[45 ] Straka L and Heczko O 2003 *J. Appl. Phys.* **93** (10) 8636

[46] Polesya S, Mankovsky S, Sipr O, Meindl W, Strunk C and Ebert H 2010 *Phys. Rev.* B 82 214409

[47] Gruner M E, Entel P, Minar J, Polesya S, Mankovsky S and Ebert H 2012 *J. Magn. Magn. Mater.* DOI: 10.1016/j.jmmm.2012.02.081

[48] Tickle R and James R D 1999 *J. Magn. Magn. Mater.* **195**, 627

[49] Buschow K H J and de Boer F R 2004 *Physics of Magnetism and Magnetic Materials* vol 97 (New York: Kluwer)

[50] Kakeshita T, Fukuda T and Takeuchi T 2006 *Mater. Sci. Eng. A* **438-440** 12

[51] Klaer P, Eichhorn T, Jakob G and Elmers H J 2011 *Phys. Rev. B* **83**, 214419

[52] Callen E R and Callen H B 1960 *J. Phys. Chem. Solids* **16**, 310

[53] Herzer G 1992 *J. Magn. Magn. Mater.* **112** 258

[54] Claussen I, Brand R, Hahn H, Mayr S G 2012 *Scripta Mater.* **66** 163

[55] Barabash S V, Chepulskii R V, Blum V, Zunger A 2009 *Phys. Rev. B* **80**, 220201(R)

[56] Chepulskii R V, Barabash S V and Zunger A, 2012 *Phys. Rev. B* **85** 144201

[57] Ren X and Otsuka K 1997 *Nature* **389** 579